\documentclass[twocolumn]{jpsj2}

\title{Collision Dynamics of Two Barchan Dunes Simulated Using a Simple Model}

\author{Atsunari \textsc{Katsuki}$^{1}$
$^{2}$\thanks{E-mail address:katsuki@cp.cmc.osaka-u.ac.jp},
Hiraku \textsc{Nishimori}$^{3}$, 
Noritaka \textsc{Endo}$^{4}$  and
Keisuke \textsc{Taniguchi}$^{4}$}

\inst{$^{1}$ Department of Physics, Osaka University, Osaka 560-0043 \\
$^{2}$ Cybermedia Center, Osaka University, Osaka 560-0043 \\
$^{3}$ Department of Mathematical Sciences, Osaka Prefecture University,
Sakai 599-8531 \\
$^{4}$ Department of Earth and Space Sciences, Osaka University, Osaka 560-0043}

\abst{
The collision processes of two crescentic dunes called barchans are
systematically studied using a simple computer simulation model.
The simulated processes, coalescence, ejection and reorganization, 
qualitatively
correspond to those observed in a water tank experiment.
Moreover we found the realized types of collision depend both on the mass 
ratio and on the lateral distance between barchans under initial
conditions.
A simple set of differential equations to describe the collision
of one-dimensional (1D) dunes is introduced.
}

\kword{sand dune, barchan dune, collision process, water tank
experiment, saltation, avalanche, numerical simulation}

\begin{document}
\maketitle
Many types of sand dunes are found in deserts, 
on the sea bottom and even on Mars.
Dunes are formed by interactions
between the flow of wind or water and sand\cite{bag1941}.
 The flow makes the shape of a dune by transporting sand particles.
The wind transports sand particles 
and forms the shape of dunes.
The dune topography, in turn, acts
as a boundary condition on
the air flow.
One of the most intensively studied types of sand dune is
the crescentic dune called the barchan.
A steadily blowing unidirectional wind generates barchans when the amount of
available sand is insufficient for covering the entire
bedrock\cite{wasson1983}.
Barchans usually migrate as a group, interacting with 
one another through collisions and interdune sand
flow\cite{bag1941,nishi1993,werner1995,nishi1998,
Gay1999,momiji2000,lima2002,
schw2003,shibata2003,her2004}.
Previous studies, however, have focused mainly on single barchans
\cite{finkel1959,wipp1986,hesp1998,sau2000,andre2002-1,kroy2002-2,sau2003}.
One of the reasons is that, because of 
the slow time scale of the system,
it is hard to observe the whole process of the interaction dynamics of barchans
in a desert.
Recently, some underwater experiments have succeeded in creating
barchans \cite{nino1997,her2002,endo2004,endo2005}, which has enabled real-time observation.
In this letter, taking only dominant factors into account,  
we propose an effective model to simulate the coaxial and offset collisions of 
two barchans, which are observed in the water tank
experiments\cite{endo2005}.

    In the model,
the dune field  is divided into square
cells\cite{nishi1993,werner1995,nishi1998}.
Each cell is considered to represent an 
area of the ground which is sufficiently larger than the sand
grains. 
A field variable $h(x,y,t)$ which expresses the local surface height
is assigned to each cell; $t$ denotes the discrete time step and
the spatial coordinates $x$ and $y$ denote the positions  of the center of a cell in the flow and the lateral directions, respectively.
The edge length of the cell is taken as a unit of length.
In short,  $x$, $y$ and $t$ are 
discrete variables while $h(x,y,t)$ takes a
continuous value.
This model belongs to a class of
simulation models called cell dynamics\cite{oono1988}.
\begin{figure}[t]
\begin{center}
\includegraphics[scale=1.]{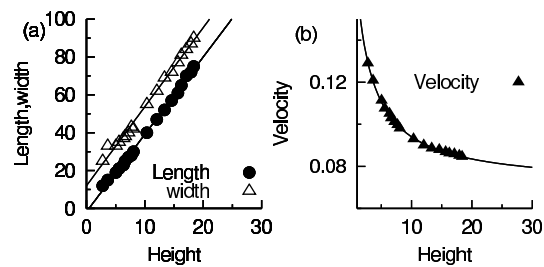}
\caption{\label{fig:HW}(a) Relationships between 
height, length and width of barchans.
The solid lines are the best linear fit.
(b) Relationships between 
barchan velocity(V) and barchan height(H). The solid line is the best fit
using the relation $V=Q/(H+H_c)+V_c$, where $H_c$ and $V_c$ are phenomenological
 parameters \cite{momiji2002}.}
\end{center}
\end{figure}
\begin{figure*}[t]
\begin{center}
\includegraphics[scale=0.6]{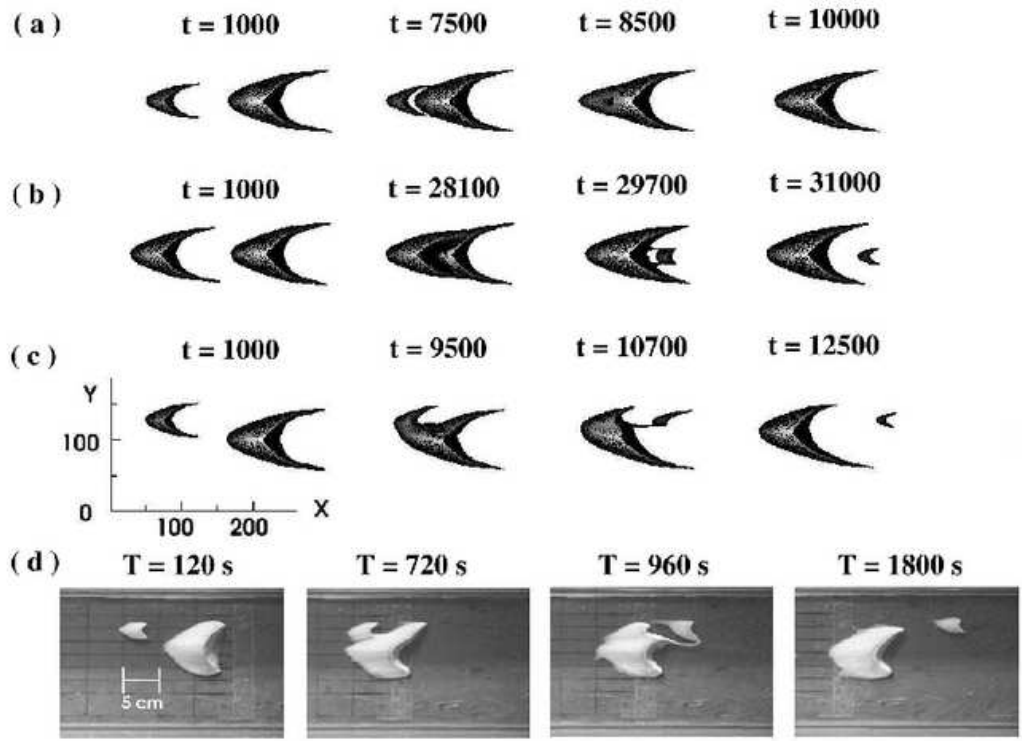}
\caption{Typical collisions of  two 
barchans: (a) coalescence process simulated with offset parameter
$\alpha=0.0$ and mass ratio $\beta=0.2$, 
 (b) ejection process simulated with offset parameter
$\alpha=0.0$ and mass ratio $\beta=0.7$,
 (c) reorganization process simulated $\alpha=0.3$ and
$\beta=0.2$, (d) offset
collision performed in water tank  which is 10 $m$ long, 20 $cm$
wide and 50 $cm$ deep with water depth maintained at approximately 13 $cm$\cite{endo2005}.
In (a), (b) and (c) the slopes at the angle of repose are painted in
 black.}
\label{fig:heighttimeg}
\end{center}
\end{figure*}
The motion of sand grains is realized by  
two processes: saltation and avalanche.
Saltation is the transportation process of sand grains
 by flow.
The saltation length and saltation mass are denoted  $L$ and $q$, respectively. Saltation occurs only for
 cells on the upwind face of dunes.
In each time step of the simulation, the amount of sand $q$ is  shifted  
from a cell  $(x,y)$ to the leeward cell $(x+L,y)$, which is the numerical  
expression of a saltation
process. 
Hence, the changes in height, $h(x,y) \rightarrow  h(x,y)-q$ and $h(x+L,y)
\rightarrow  h(x+L,y)+q$, take place at the taking-off cell and the 
landing cell, respectively.
The saltation length $L$ and
the amount of transported sand $q$  
are modeled by the following rules,
\begin{eqnarray}
L &=& a+bh(x,y,t)-ch^2(x,y,t),\label{eq:L}\\
q &=& d,
\label{eq:Ql}
\end{eqnarray}
where  a=1.0, b=1.0, c=0.01 and d=0.1 are phenomenological parameters.
In  eq. (\ref{eq:L}), the second term shows that sand is transported
farther away
as the height of the sand surface is higher. The last term is introduced for
$L$ not to become too large.
Note that (\ref{eq:L}) is used only in the range where $L$ increases as a function of $h(x,y,t)$.
The saltation mass $q$ is fixed at 0.1 for simplicity.
In the avalanche process, on the other hand, 
the sand grains slide down along the locally steepest slope until the slope relaxes to be (or be lower than) the angle of repose which is set 
to be $34^{\circ}$.

We examine if some basic features of barchan are realized by this simple
model
\cite{finkel1959,hesp1998,sau2000}.
Starting from an initial Gaussian sand pile,
we measured the morphologic relations
after a steady barchan shape was reached.
The linear 
relationships between height, length (Fig. \ref{fig:HW}(a)) and width(Fig. \ref{fig:HW}(b)) are shown.
Next, the roughly inverse relationship between migration velocity 
and height(Fig. \ref{fig:HW}(c)) was confirmed to hold. 
In these tests, quasi-periodic
boundary conditions were used in which the total mass of sand  
flowing away from the downwind 
and lateral boundaries 
was homogeneously re-injected from
the upwind boundary.

In order to simulate  collision processes,
two initial barchans are situated in a field. 
The longitudinal ($\it i.e,$ wind directional) distance 
between their crests 
is set to $d_x$ =2$L_l$, where $L_l$ is the main body length of the initial 
leeward barchan, while the lateral distance between them 
is set to $d_y$=$\alpha W_l$, where $W_l$ is the width of 
the same leeward barchan. 
The constant $\alpha$ is varied as a control parameter which we call
the $\it offset$ $\it parameter$.
We fix the initial mass of the leeward barchan ($M_l$=10) and
define its ratio ($\beta \equiv M_w / M_l$) 
to the initial mass of the windward barchan ($M_w$) as another
control parameter.  

In the first simulation, we examine the coaxial collisions.
For the sake of quantitative comparison with experiments,
open boundary conditions are used.
This means that there is no influx sand from the upwind boundary and the sand out of the numerical field from
 the downwind boundary is neglected.
Figure \ref{fig:heighttimeg}(a) shows the simulated process 
for $\alpha$ = 0.0 and $\beta$ = 0.2.
After two barchans collide with each other,        
the windward barchan is absorbed into the leeward barchan,
during which the upwind slope of the leeward barchan is temporally
eroded and forms a dent.
As time proceeds, the dent is filled up and the unified dune recovers the crescent shape of the barchan.
The behavior of barchans which is  similar to this simulation 
result was observed in water tank experiments \cite{endo2005}.
Hereafter, we call such a process the $\it coalescence $ of two barchans. 
Figure \ref{fig:heighttimeg}(b) shows the time evolution for 
 $\alpha=0.0$ and $\beta = 0.7$.
Similar to the case of coalescence,
in the initial stage, a part of the windward barchan merges
into the leeward barchan.  However, the valley between
two crests is kept imperfectly filled, thereafter, a small barchan is 
$\it bled$ from the
downwind slope.
\begin{table}[t]
 \caption{Calculated results classified into
  coalescence ($\times$) , ejection ($\bigtriangleup$) and
 reorganization ($\circ$) for various mass ratios
and offset parameters. 
}
 \label{table:mass-off-set}
\begin{center}
\begin{tabular}{c c c c c c c c} 
 Offset &  \multicolumn{6}{c}{Mass ratio($\beta$)} \\ 
 parameter ($\alpha$) & 0.1 & 0.2 & 0.3 & 0.4 & 0.5 & 0.6 & 0.7   
 \\ 
\hline
      0 & $\times$ & $\times$ & $\times$ & $\times$ & $\times$ 
            & $\bigtriangleup$  & $\bigtriangleup$   \\
   0.1 & $\times$ & $\times$ & $\circ$  & $\circ$  & $\circ$
            & $\circ$  & $\circ$    \\  
    0.2 & $\times$ & $\circ$  & $\circ$  & $\circ$  & $\circ$
            & $\circ$  & $\circ$   \\             
    0.3 & $\circ$  & $\circ$  & $\circ$  & $\circ$  & $\circ$ 
            & $\circ$  & $\circ$   \\
    0.4 & $\circ$  & $\circ$  & $\circ$  & $\circ$  & $\circ$      
            & $\circ$  & $\circ$   \\
    0.5 & $\circ$  & $\circ$  & $\circ$  & $\circ$  & $\circ$     
            & $\circ$  & $\circ$    
  \end{tabular}
\end{center}
\end{table}
\begin{figure}
\begin{center}
\includegraphics[scale=1.]{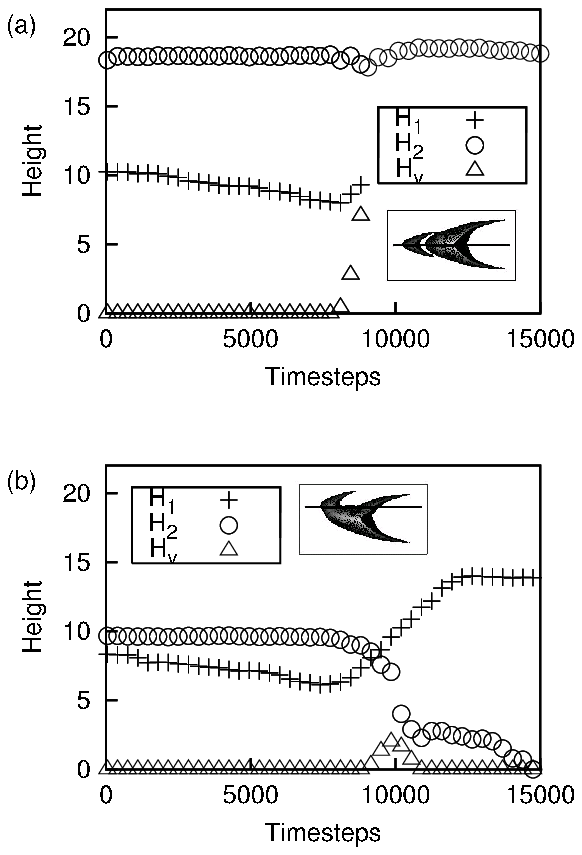}
\caption{
\label{fig:htg}Time evolution of heights 
$ \textit{\textbf{H}} \equiv (H_1, H_2,H_v)$
in longitudinal profile
 along solid line as indicated in inset.
 (a) For coaxial collision (Fig. 2(a))
the coalescence of two crests is seen at 
approximately 9000 time steps.
 (b) For offset collision (Fig. 2(c))
the exchange of the heights of two crests is 
seen at approximately 9200 time steps,
 which indicates the reorganization of two barchans.}
\end{center}
\end{figure}
Although the small barchan firstly drags ridges from its horns 
connecting
to the windward barchan, the ridges are soon cut and finally steady 
shapes of windward and leeward barchans are attained.
Hereafter, we call such a process the $\it ejection $ of a small barchan.
Figure \ref{fig:heighttimeg}(c) shows the time evolution for 
 $\alpha=0.3$ and $\beta = 0.2$.
As the initial windward barchan approaches the leeward barchan, the wing of the leeward barchan is partially eroded 
and is pushed downward.
After that it is separated from the deforming main body.
The separated body
drags a ridge extending from the main body for a while. 
Subsequently, it is completely isolated and recovers the  crescent shape. Hereafter we call such a process the $\it reorganization $ of two barchans. 
This process is qualitatively similar to what
happened in the experimental tank (Fig. \ref{fig:heighttimeg}(d)).
Note that the types of realized collision 
depend on the mass ratios 
of initial barchans 
and on their relative positions.

In addition,  systematic calculations are performed  
with various combinations of $\alpha$ and $\beta$. 
The obtained collisions are classified into three types,
coalescence, ejection and reorganization (Table \ref{table:mass-off-set}). 
Note that the coalescence appears in the region of
low values of offset parameter and small mass ratios,
and the reorganization occurs at high values of offset
parameter or large mass ratios. 
On the other hand, ejection occurs at low values of offset
parameter and large mass ratios, which corresponds 
to the ``solitary-wave
behavior'' obtained by the previous simulation \cite{schw2003}.
This diagram should be compared with those of 
further subaqueous experiments and also to those of 
field observations.

To explore the intrinsic mechanism underlying 
the difference between the three types of 
collision, we focus on 
the dynamics of longitudinal profiles of barchans particularly on the 
heights of the windward crest ($H_1$), the leeward 
crest ($H_2$) and 
the bottom of the valley ($H_v$).
Figure $\ref{fig:htg}$(a) shows the time evolution of 
$ \textit{\textbf{H}} \equiv (H_1, H_2,H_v)$
 in the
case of coalescence. $H_v$ reaches $H_1$ at 
approximately 9000 time steps.
Figure $\ref{fig:htg}$(b) shows the time evolution of 
$\textit{\textbf{H}}$
 in the case of 
 reorganization. The exchange of 
the heights of two barchans
is seen at approximately 9200 time steps,
whereas the bottom 
height of the valley is kept  
below them throughout the process.
The time evolution in the case of ejection is qualitatively the same as
that in the case of reorganization.
In short, ejection is a type of reorganization.
We discuss only coalescence and reorganization hereafter.

The dynamics in each profile of Fig. $\ref{fig:htg}$ is not self-complete
because of the lateral sand flow. 
Nevertheless, the dynamics realized in each profile
is roughly inferred  from the initial condition,  that is,
(i) if the height ratio of the windward crest to the leeward 
crest
is comparatively small under the initial condition,  coalescence occurs.
(ii) if the height ratio is comparatively large, 
the reorganization of two barchans occurs.
In process (i), the smaller barchan on the windward side 
climbs the upward face 
of the
larger barchan on the leeward side 
and is absorbed on the latter before its crest 
reaches the
same height as the latter.  On the other hand, in process (ii), the 
windward barchan
climbs the larger leeward barchan and its crest 
becomes higher than that of the 
leeward barchan
without absorption into the latter.
Subsequently, the leeward barchan 
runs away from the windward barchan
because of the inverse relationship between velocity and height.
From the above discussion, the important factor determining 
the type of collision is presumed to be the competition between 
the absorption and height exchange of 
two barchans.  
\begin{figure}
\begin{center}
\includegraphics[scale=0.8]{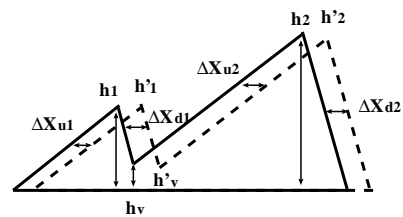}
\caption{\label{fig:after}Schematic diagram for collision process of 1D barchan collision.  
The solid and dashed lines represent
 the profiles at time $t$ and time $t+\Delta t$, respectively.}
\end{center}
\end{figure}

To attain a clear picture of the complex processes of collisions,
 we further simplify the model using a set of naive assumptions.
The first assumption is the geometric similarity 
between 1D dunes of different sizes which means
 that constant angles of upwind and 
downwind slopes are maintained irrespective of their size. 
The second assumption is that the sand flux at the crests of dunes is
 described by 
the product of eqs. (\ref{eq:L}) and (\ref{eq:Ql}).
The third assumption is the mass conservation of sand crossing over
 a crest, and that all the sand blown across a 
crest becomes trapped in the downwind face of the same dune without 
directly escaping to the leeward. 
The combination of the above assumptions leads to a simple conservation relationship between the eroded sand on the upwind faces  
and the accumulated sand on the downwind faces of each
 dune(Fig. $\ref{fig:after}$). 
The sand mass conservation in the 
windward dune is expressed as,         
\begin{equation}   
 \rho \frac{(h_1+ h_1')}{2} \Delta x_{u1} =
 \rho \frac{(h_1+ h_1')-(h_v+ h_v' )}{2} \Delta x_{d1} \equiv q_1\Delta t,
 \label{eq:B1}
\end{equation}
where $h_1$ and $h_v$ are the heights of the windward dune and the 
bottom of the valley at time $t$, respectively, whereas  
$h_1'$ and $h_v'$ indicate the heights of the windward dune and the bottom of the valley at time $t+\Delta t$.
$\Delta x_{u1}$ and $\Delta x_{d1}$ denote the horizontal displacements 
of the upwind surface and the downwind surface of the windward dune during $\Delta t$,
$q_1 \equiv q(h_1)$ denotes the sand mass crossing over the crest per unit time, and $\rho$ denotes
the area density of the bulk of sand. 
Also,
the change in height of the windward crest for
 $\Delta t$
is expressed as, 
\begin{equation}
 h_1'-h_1  = A\Delta x_{d1}-A\Delta x_{u1}\equiv \Delta h_1,
 \label{eq:B3}
\end{equation}
where $A$ is the geometrical constant of similar triangles 
constituting 1D dunes, namely, the ratio of their heights to the lengths of their bases.
Using eqs. (\ref{eq:B1}) and (\ref{eq:B3}) and taking the limits $\Delta t \rightarrow 0$, 
$\Delta x_{u1} \rightarrow 0$, $\Delta x_{d1} \rightarrow 0$ and $\Delta h_1 \rightarrow 0$,
the evolution equation of the crest height, 
\begin{equation}
\frac{dh_1}{dt}=\frac{q_1 A}{\rho} (\frac{1}{h_1-h_v}-\frac{1}{h_1}),
\label{eq:B5}
\end{equation}
is obtained.
Similarly the evolution equations of the heights of the leeward dune and the
valley are written as,
\begin{eqnarray}
\frac{dh_2}{dt}&=&\frac{q_2 A}{\rho} (\frac{1}{h_2}-\frac{1}{h_2-h_v}),\label{eq:B6} \\
\frac{dh_v}{dt}&=&\frac{q_1 A}{\rho} (\frac{1}{h_1-h_v})-\frac{q_2 A}{\rho}(\frac{1}{h_2-h_v}).
\label{eq:B7}
\end{eqnarray}
\begin{figure}
\includegraphics[scale=1.]{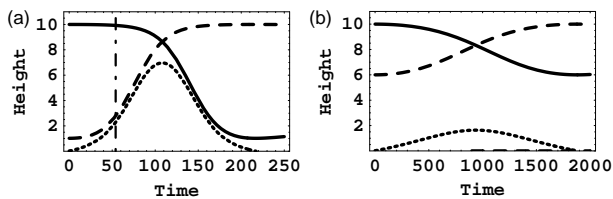}
	\caption{\label{fig:1d-mathe-g}Time evolution 
of heights of windward dune(dashed), leeward dune(solid) and valley(dotted) obtained by 
 numerical calculation. The dash-dotted line represents time when $\delta h$ reaches
 $\delta h_{min}$ = 0.5. 
(a) Under the initial conditions of $h_1(0)=1.02$,
$h_2(0)=10.0$ and $h_v(0)=0.01$, 
coalescence collision occurs.
(b) Under the initial conditions of $h_1(0)=6.00$,
$h_2(0)=10.0$ and 
$h_v(0)=0.01$, the reorganization occurs.}
\end{figure}

In addition to the above closed set of differential 
equations, we set the minimum 
depth as $\delta h_{min}$.
Here, the valley depth $\delta h$ is defined as 
$\delta h = h_1 -h_v$ and two dunes are considered to coalesce
once $\delta h$ reaches $\delta h_{min}$.
Figure $\ref{fig:1d-mathe-g}$(a) shows a typical collision obtained by
the calculation of
eqs. (\ref{eq:B5}), (\ref{eq:B6}) and (\ref{eq:B7}) with $\delta h_{min}=0.5 $. 
In this case, the coalescence of two dunes occurs at the time represented by the 
dash-dotted line. $h_v$ reaches $h_1$
before the exchange of heights occurs and coalescence occurs.
This process is similar to that indicated in Fig. $\ref{fig:htg}$(a).
Figure $\ref{fig:1d-mathe-g}$(b) shows another typical collision 
under the same initial conditions as indicated 
in Fig. $\ref{fig:1d-mathe-g}$(a) except for the 
higher windward initial dune.
The exchange of heights occurs before $h_v$ reaches
$h_1$ and reorganization occurs.
This process is similar to that indicated in Fig. $\ref{fig:htg}$(b).
This indicates that the set of simple differential 
equations, eqs. (\ref{eq:B5}), (\ref{eq:B6}) and (\ref{eq:B7}), contains the intrinsic features of collision dynamics.

To summarize,  
the coaxial and offset collisions of two barchans are simulated using a simple model that includes only saltation and avalanche processes without
taking complex wind flow into account. 
Also we introduced a set of differential 
equations, which effectively describe the collision processes of 1D dunes.

We thank Macoto Kikuchi for helpful discussions. 
This work was partially supported by 
the 21st Century COE Program, "Towards a new basic science : 
 depth and synthesis" and
 the Grant-in-Aid for Scientific Research (C) 
No. 16605008 of
the Ministry of Education, Culture, Sports, Science and Technology.


\begin{thebibliography}{9}
\bibitem{bag1941}R. A. Bagnold : \textit{The Physics of Blown Sand and
Desert Dunes}  (Methuen, London, 1941).
\bibitem{wasson1983}R. J. Wasson and R. Hyde
	:Nature \textbf{304} (1983) 337
\bibitem{nishi1993}H. Nishimori and N. Ouchi
	:Phys. Rev. Lett. \textbf{71} (1993) 197.
\bibitem{werner1995}B. T. Werner
	:Geology \textbf{23} (1995) 1107.
\bibitem{nishi1998}H. Nishimori, M. Yamasaki and K. H.  Andersen
	:J. Mod. Phys. B \textbf{12} (1998) 256.
\bibitem{Gay1999} S. P. Gay: Geomorphology  \textbf{27} (1999) 279.
\bibitem{momiji2000}H. Momiji and R. Carretero-Gonzalez and S. R. Bishop and A. Warren
	:Earth Surface Processes and Landforms \textbf{25} (2000) 905.
\bibitem{lima2002} A. R. Lima,  G. Sauermann, H. J. Herrmann and K. Kroy
	:Physica A \textbf{310} (2002) 487. 
\bibitem{schw2003} V. Schw$\ddot{a}$mmle and H. J. Herrmann
	:Nature \textbf{426} (2003) 619.
\bibitem{shibata2003} J. Shibata and Y-h. Taguchi
	:J. Phys. Soc. Jpn. \textbf{72} (2003) 2685.
\bibitem{her2004} P. Hersen, K. H. Andersen, H. Elbelrhiti,
	B. Andreotti, P. Claudin and S. Douady
	:Phys. Rev. E \textbf{69} (2004) 011304.
\bibitem{finkel1959} H. J. Finkel: Journal of Geology  \textbf{67}
	(1959) 614.
\bibitem{wipp1986}F. K. Wippermann and G. Gross
	:Boundary-Layer Meteorology \textbf{36} (1986) 319.
\bibitem{hesp1998} P. A. Hesp  and K. Hastings
	:Geomorphology \textbf{22} (1998) 193.
\bibitem{sau2000}G. Sauermann, P. Rognon, A. Poliakov and H. J. Herrmann
	:Geomorphology \textbf{36} (2000) 47.
 \bibitem{andre2002-1}B. Andreotti, P. Claudin and S. Douady
	:Eur. Phys. J. B \textbf{28} (2002) 321.
\bibitem{kroy2002-2}K. Kroy, G. Sauermann and H. J. Herrmann
	:Phys. Rev. E \textbf{66} (2002) 031302.
\bibitem{sau2003}G. Sauermann , J. S. Andrade Jr., L. P. Maia, U. M. S. Costa,
	A. D. Araujo and H. J. Herrmann
	:Geomorphology \textbf{1325} (2003) 1.
\bibitem{nino1997} Y. Ni$\tilde{n}$o and  M. Barahona
	:Proc. Int. Assoc. Hydraul. Res. \textbf{27B} (1997) 1037. 
\bibitem{her2002} P. Hersen,  S. Douady and B. Andreotti
	:Phys. Rev. Lett. \textbf{89} (2002) 264301. 
\bibitem{endo2004} N. Endo, H. Kubo and T. Sunamura: Earth
	Surface Processes and Landforms \textbf{29} (2004) 31.
\bibitem{endo2005}N. Endo, K. Taniguchi and A. Katsuki
	:Geophysical Research Letters \textbf{31} (2004) 12503.
\bibitem{oono1988}Y. Oono and S. Puri
	:Phys. Rev. A  \textbf{38} (1988) 434.
\bibitem{momiji2002} H. Momiji and H. Nishimori and S. R. Bishop
	:Earth Surface Processes and Landforms \textbf{27} (2002) 1335.
\end{thebibliography}
\end{document}